\documentclass[11pt,a4paper]{article}
\usepackage{jheppub}
\usepackage{setspace,braket}
\usepackage{amsmath, amssymb, bm}
\usepackage{etoolbox}

    \makeatletter
    \patchcmd{\maketitle}{\@fpheader}{}{}{}
\makeatother
\def\be{\begin{equation}}
\def\ee{\end{equation}}

\def\({\left(}
\def\){\right)}
\def\[{\left[}
\def\]{\right]}

\newcommand{\bea}{\begin{eqnarray}}
\newcommand{\eea}{\end{eqnarray}}

\def\d#1#2{\frac{\displaystyle #1}{\displaystyle #2}}

\numberwithin{equation}{section}
\renewcommand{\thesection}{\arabic{section}}

\begin{document}
\renewcommand{\thefootnote}{\fnsymbol{footnote}}

\title{Noncommutative geometry inspired black holes in Rastall gravity}
\author[a,b]{Meng-Sen Ma,}
\author[a]{Ren Zhao}
\affiliation[a]{Institute of Theoretical Physics, Shanxi Datong
University, Datong 037009, China}
\affiliation[b]{Department of Physics, Shanxi Datong
University, Datong 037009, China}
\emailAdd{mengsenma@gmail.com; ms\_ma@sxdtdx.edu.cn; zhao2969@sina.com}

\abstract{

Under two  different metric ansatz, the noncommutative geometry inspired black holes (NCBH) in the framework of Rastall gravity are derived and analyzed. We consider the fluid-type matter with the Gaussian-distribution smeared mass density. Taking a Schwarzschild-like metric ansatz, it is shown that the noncommutative geometry inspired Schwarzschild black hole (NCSBH) in Rastall gravity, unlike its counterpart in general relativity (GR), is not a regular black hole. It has at most one event horizon. After undergoing a finite maximal temperature, the black hole will leave behind a point-like massive remnant at zero temperature. Considering a more general metric ansatz and a special equation of state of the matter, we also find a regular NCBH in Rastall gravity, which has the similar geometric structure and temperature to that of NCSBH in GR.

}
\maketitle
\onehalfspace

%%%%%%%%%%%%%%%%%%%%%%%%%%%%%%%%%%%%%%%
%%%%%%%%%%%%%%%%%%%%%%%%%%%%%%%%%%%%%%%
%%%%%%%%%INTRODUCTION%%%%%%%%%%%%%%%%%%%
%%%%%%%%%%%%%%%%%%%%%%%%%%%%%%%%%%%%%%%
\renewcommand{\thefootnote}{\arabic{footnote}}
\setcounter{footnote}{0}
\section{Introduction}
\label{intro}

General relativity (GR), which has many successful predictions, is the most popular theory of gravity. The Einstein field equation can be derived via the variational principle from a total action (the gravitational action plus the matter action). Due to the Bianchi identity, the covariant conservation of the energy-momentum tensor, namely $T^{\mu\nu}_{~~;\mu}=0$,  is naturally satisfied. In GR and many other theories of gravity, it is assumed that the geometry and the matter fields are coupled to each other in a minimal way. It has been shown that when the geometry and the matter fields are coupled in a non-minimal way, such as the direct curvature-matter coupling, the covariant conservation of the energy-momentum tensor may be violated\cite{Kremer.023503.2004,Nojiri.137.2004,Balakin.1867.2005,Koivisto.4289.2006,Bertolami.104016.2007,Balakin.084013.2008,Sotiriou.225.2008,Harko.024020.2011,Harko.410.2014}. In fact, the covariant conservation of the energy-momentum tensor is just an assumption and has not been generally tested by observation. Thus, by relaxing the condition of covariant conservation of the energy-momentum tensor, Rastall proposed a phenomenological gravitational model by considering $T^{\mu\nu}_{~~;\mu} \propto R^{;\nu}$ with $R$ the Ricci scalar\cite{Rastall.3357.1972}. Rastall gravity has been employed to study the cosmological consequences\cite{AlRawaf.935.1996,Batista.593.2012,Moradpour.259.2017}. Various exact solutions of Rastall gravity have also been derived in \cite{Carames.3145.2014,Mello.085009.2015,Oliveira.044020.2015,Bronnikov.162.2016,Oliveira.124020.2016}.

In Rastall gravity, the energy-momentum tensor can also be derived from the Lagrangian of matter fields, namely $T_{\mu\nu}=-2\frac{\delta\mathcal{L}_m}{\delta g^{\mu\nu}}+g_{\mu\nu}\mathcal{L}_m$. However, there is no the  corresponding equation of motion for the matter fields due to the lack of a total action for Rastall gravity. For fluid-type matter, it is relatively simple to find the exact solution in Rastall gravity because in this case one does not  need to consider the equation of motion of the matter fields, but only the equation of state. In this paper, we will consider this kind of matter.

By considering the noncommutativity of spacetime, Nicolini, et.al studied several kinds of noncommutative geometry inspired black holes (NCBH) in GR\cite{Nicolini.547.2006,Ansoldi.261.2007,Nicolini.1229.2009,Spallucci.449.2009,Nicolini.015010.2010}. Noncommutativity eliminates point-like structures in favor of smeared objects. The conventional mass density of point-like source can be replaced by a smeared, Gaussian distribution:
\be\label{rho}
\rho =\frac{M }{(4 \pi  \theta )^{3/2}}\exp \left(-\frac{r^2}{4 \theta }\right),
\ee
where $\theta$ is a constant with dimension of length squared and represents the noncommutativity of spacetime.
The effect of noncommutative geometry on gravity has been contained in the matter source. As stated in \cite{Nicolini.547.2006}, ``the noncommutativity is an intrinsic property of the manifold itself, rather than a super-imposed
geometrical structure", so noncommutativity can also be considered in other theories of gravity, besides GR. Due to different gravitational field equations, we expect that the NCBH in modified gravities may have some different properties. For example, the currently known NCBH in GR are all regular black holes because the singularity at the origin is usually replaced by a de Sitter core. We want to know whether this is also true for other theories of gravity. Among the numerous modified gravities, Rastall gravity has simpler field equations and thus can be easily dealt with. In this paper, we will study the NCBH in Rastall gravity.

The plan of this paper is as follows:
In Sec.2 we simply introduce the Rastall gravity.
In Sec.3 We first present the spherically symmetric, static vacuum solution with point-like source in Rastall gravity and then study the NCBH in Rastall gravity.
In Sec.4 we summarize our results and discuss the possible future directions.

\section{Rastall gravity}

The field equations for Rastall gravity read
\bea
&&G_{\mu\nu}+\kappa\lambda g_{\mu\nu}R=\kappa T_{\mu\nu},\label{Rastall}\\
&&T^{\mu\nu}_{~~;\mu} =\lambda R^{;\nu},\label{emt}
\eea
where $\lambda$ is the Rastall parameter and $\kappa$ is the Rastall gravitational coupling constant. The corresponding trace equation is given by
\be\label{trace}
R=\d{\kappa}{4\kappa\lambda-1}T,
\ee
where $T=T^\mu_{~\mu}$ is the trace of the energy-momentum tensor and $R$ is the Ricci scalar. Employing the trace equation, Eq.(\ref{Rastall}) and Eq.(\ref{emt}) can also be written into
\bea
&&G_{\mu\nu}=\kappa\left(T_{\mu\nu}-\d{\kappa\lambda}{4\kappa\lambda-1}g_{\mu\nu}T\right),\\
&&T^{\mu\nu}_{~~;\mu} =\d{\kappa\lambda}{4\kappa\lambda-1} T^{;\nu}.\label{emtt}
\eea
When $\lambda=0$, the traditional GR and the covariant conservation of the energy-momentum tensor are both recovered with the parameter $\kappa=8\pi G$.

To recover the Newtonian gravity in the weak-field approximation, the parameters in Rastall gravity should satisfy the relation\cite{Rastall.3357.1972}:
\be
\d{\kappa(6\kappa\lambda-1)}{4\kappa\lambda-1}=8\pi G.
\ee
Therefore, $\kappa\lambda = 1/4$ and $\kappa\lambda = 1/6$ are not allowed in principle.

Below we will consider a special case with $\kappa\lambda = 1/2$ and correspondingly $\kappa=4\pi G$.
In this case, the gravitational field equation turns into
\be\label{eom}
R_{\mu\nu}=4\pi G T_{\mu\nu},\; \text{or}\;\; G_{\mu\nu}=4\pi G\left(T_{\mu\nu}-\d{1}{2}g_{\mu\nu}T\right),
\ee
and Eq.(\ref{emtt}) becomes
\be\label{eom2}
T^{\mu\nu}_{~~;\mu} =\d{1}{2} T^{;\nu}.
\ee

In this paper, we will consider a fluid-type matter source with the energy-momentum tensor $T^\mu_{~\nu}$ satisfying:
\be
T^\mu_{~\nu}=\text{Diag}\left(-\rho(r),p_r(r),p_\perp(r),p_\perp(r)\right),
\ee
we try to find the static, spherically symmetric solution of Rastall gravity.

As a warmup, we first consider the gravitational field of a point-like source with mass $M$ located at the origin. The mass density is
\be
\rho(r)=-T^t_{~t}=\d{M}{4\pi r^2}\delta(r).
\ee
For the metric ansatz,
\be\label{metric1}
ds^2 =-f(r) dt^2 +\frac{dr^2}{f(r)} +r^2
d\Omega^2,
\ee
the nonzero components of $R^\mu_{~\nu}$ are
\bea
R^t_{~t}&=&R^r_{~r}=-\frac{f''(r)}{2}-\frac{f'(r)}{r};\\
R^\theta_{~\theta}&=&R^\phi_{~\phi}=-\frac{r f'(r)+f(r)-1}{r^2}.
\eea
According to Eq.(\ref{eom}), we can easily obtain the expression of $f(r)$:
\be\label{noNC}
f(r)=C_1-\d{C_0+2MG}{r}.
\ee
If requiring the metric approaches to the Minkowski space at infinity, the integration constant $C_1$ should be $C_1=1$. Furthermore, if removing the source (namely $M=0$), the vacuum solution must be Minkowski one.
Thus, the integration constant $C_0$ should be zero. This is just the well-known Schwarzschild solution outside the source, which is the solution in vacuum. This result is expectable. Because Rastall gravity is equivalent to GR in the vacuum case.

\section{Noncommutative geometry inspired black holes}

Considering the noncommutativity of spacetime, the matter source can be diffused throughout a region, instead of locating at a point.
For the smeared matter distribution, the mass density is given by Eq.(\ref{rho}),
\be
\rho =-T^t_{~t}=\frac{M }{(4 \pi  \theta )^{3/2}}\exp \left(-\frac{r^2}{4 \theta }\right).
\ee
Under the same metric ansatz, Eq.(\ref{metric1}), we can easily obtain
\be\label{fr1}
f(r)=B_1-\d{B_0}{r}-\d{2GM}{r\sqrt{\pi}} \gamma(1/2,r^2/4\theta),
\ee
where $\gamma(1/2,r^2/4\theta)$ is the lower incomplete gamma function with definition
  \be
  \gamma(1/2,r^2/4\theta)=\int_0^{r^2/4\theta}dt~ t^{-1/2}e^{-t}.
  \ee
The same reason as that for the point-like source in the previous section, the two integration constants should be $B_1=1$ and $B_0=0$. Thus, the metric function reads
%\footnote{A similar result is also obtained in \cite{123}, where the error function, but not lower incomplete gamma function is used. The error function is defined as $\text{erf}(x)\equiv \frac{2}{\sqrt{\pi }}\int _0^x e^{-t^2}d t$. In fact, $\gamma(1/2,r^2/4\theta)=\sqrt{\pi}\text{erf} \left(r/2\sqrt{\theta}\right)$ .}
\be\label{sol1}
f(r)=1-\d{2GM}{r\sqrt{\pi}} \gamma(1/2,r^2/4\theta).
\ee
This solution asymptotically approaches to the Schwarzschild solution due to $\gamma(1/2,r^2/4\theta) \rightarrow \sqrt{\pi}$ at infinity. This is the noncommutative geometry inspired Schwarzschild black hole (NCSBH) in Rastall gravity. 

Surprisingly, this solution is also obtained in a two-dimensional model of noncommutative spacetime\cite{Nicolini.2005}, where the linearized noncommutative version of the Einstein equation is considered. However, there is a major difference between the two black hole solutions. In two-dimensional spacetime, the black hole solution (\ref{sol1}) is regular everywhere, because the Ricci scalar $R=-\frac{GM}{3 \left(\sqrt{\pi } \theta ^{3/2}\right)}$ at the origin. We will show that in the four-dimensional case this solution is singular at the origin.
Expanding the metric function near $r=0$, it is
\be
f(r)=1-\frac{2GM}{\sqrt{\pi\theta}}+\frac{GM r^2}{6 \sqrt{\pi } \theta ^{3/2}}+O(r^4).
\ee
Due to the redundant term $\frac{2GM}{\sqrt{\pi\theta}}$, unlike the NCSBH in GR\cite{Nicolini.547.2006}, there is no de Sitter core at a short distance. Although the NCSBH in Rastall gravity is finite at the origin, it is not a  regular black hole. This can be easily found out according to the Ricci scalar, which is
\be
R=-\frac{GM e^{-\frac{r^2}{4 \theta }} \left(r^2-4 \theta \right)}{\sqrt{\pi } \theta ^{3/2} r^2}.
\ee
Clearly, it diverges in the limit $r\rightarrow 0$. In fact, for the solution Eq.(\ref{sol1}), the tangential pressure $p_\perp(r)$ is also divergent near the origin, which also reflects the singular behavior of the NCSBH at the origin.

\begin{figure}[htp]
\centering{
\includegraphics[width=7cm,keepaspectratio]{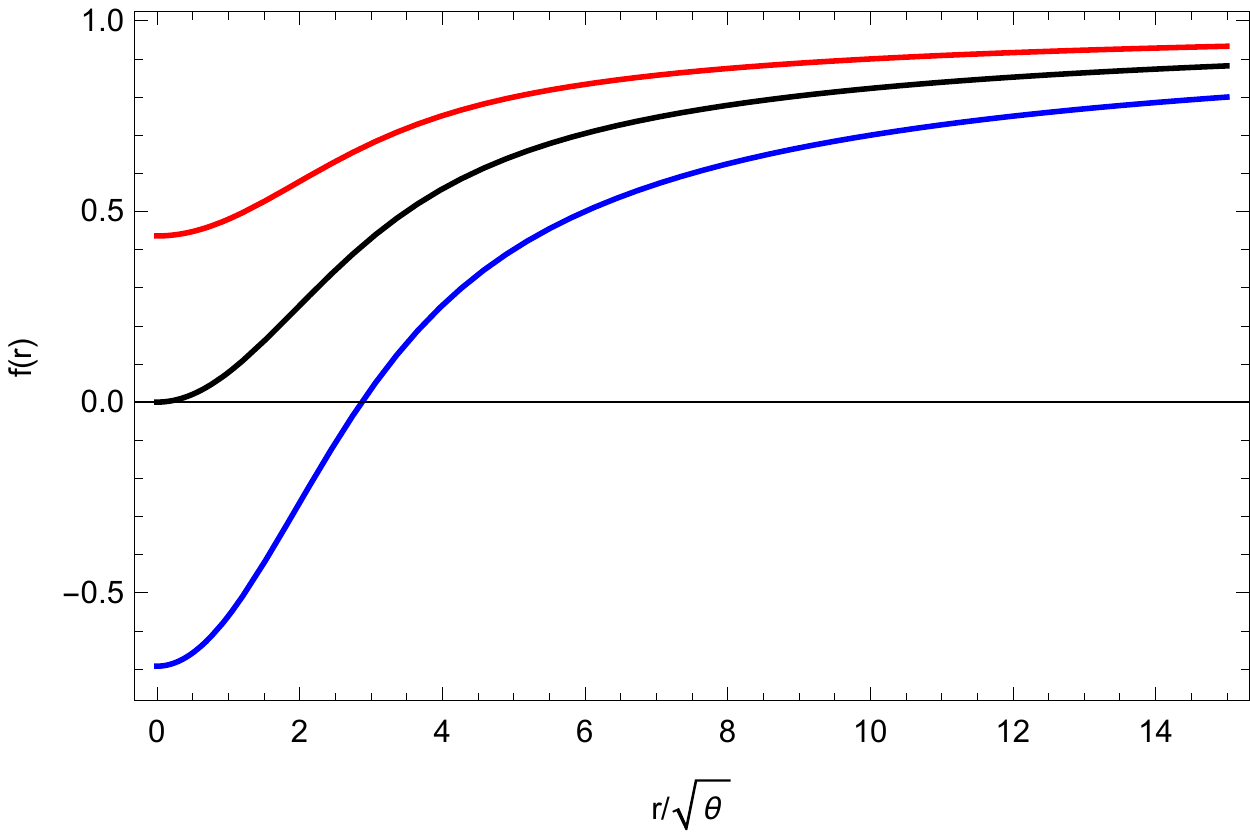}
\caption{$f(r)$ as function of $r$. The curves from top to bottom correspond to the values of $M/\sqrt{\theta}=0.5,~\sqrt{\pi}/2,~1.5$ respectively. We set $G=1$.}\label{f1}}
\end{figure}

Another difference from the NCSBH in GR is that the NCSBH in Rastall gravity has at most one event horizon. Fig.\ref{f1} shows the structure of the metric function. When $M/\sqrt{\theta}<\sqrt{\pi}/2$, the NCSBH has no event horizon and it has one event horizon when $M/\sqrt{\theta}>\sqrt{\pi}/2$. This property is special. Schwarzschild black hole always has one horizon, $r_H=2M$,  while the NCSBH in GR can have at most two event horizons.

\begin{figure}[htp]
\center{
\includegraphics[width=7cm,keepaspectratio]{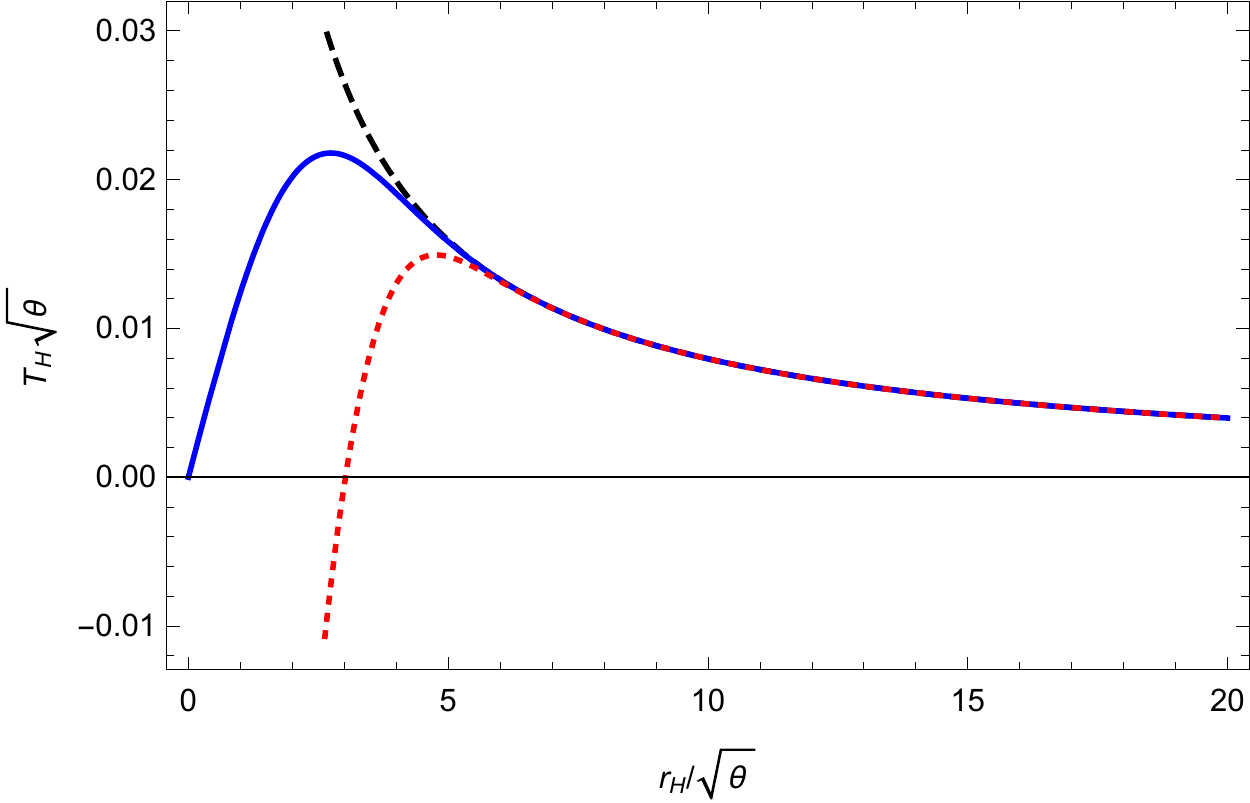}
\caption{Behaviors of the Hawking temperature $T_{H}$ for three kinds of black holes. The solid (blue) curve corresponds to the NCSBH in Rastall gravity. The dashed (black) curve is the temperature for Schwarzschild black hole. And the dotted (red) curve depicts the temperature of NCSBH in GR. }\label{T1}}
\end{figure}

The temperature of the NCSBH in Rastall gravity is given by
\be
T_H=\d{f'(r_{H})}{4\pi}=\frac{1}{4\pi r_{H}}\left[1-\frac{r_{H}e^{-\frac{r_H^2}{4 \theta }}}{\sqrt{\theta } \gamma(1/2,r_{H}^2/4\theta)}\right].
\ee
As is depicted in Fig.\ref{T1}, for larger value of $r_H/\sqrt{\theta}$ the temperature of the NCSBH in Rastall gravity is nearly coincident with those of Schwarzschild black hole and the NCSBH in GR. For smaller value of $r_H/\sqrt{\theta}$, the behavior of the temperature of the NCSBH in Rastall gravity is more similar to that of the NCSBH in GR. However, the NCSBH in GR has a zero-temperature remnant with radius about $r_H=3\sqrt{\theta}$ and the NCSBH in Rastall gravity will evaporate leaving behind a point-like massive remnant.

Several comments are needed.

(1). The thermodynamic properties of the NCBH in GR have been studied extensively\cite{Myung.012.2007,Banerjee.124035.2008,Mehdipour.124049.2010,Nicolini.097.2011,Liang.30001.2012,Ma.1750018.2017}. Here we only discuss the temperature of the NCSBH in Rastall gravity because Hawking radiation is a kinematic effect and the temperature can be derived by the metric only.
Other thermodynamic quantities, such as black hole entropy, ADM mass, depends on the Lagrangian of the gravitational theories. Whereas, for Rastall gravity there is no corresponding Lagrangian. A possible way to derive the black hole entropy only by the field equation is the method of horizon thermodynamics\cite{Padmanabhan.5387.2002}, which has been used to study the thermodynamic properties of black hole and cosmology in Rastall gravity\cite{Moradpour.187.2016,Moradpour.3492796.2016}. Many other applications of horizon thermodynamics in black hole thermodynamics can be found in \cite{Paranjape.104015.2006,Cai.084061.2010,Miao.1450093.2014,Hu.024006.2015,Ma.278.2015,Ma.351.2017}.

(2). If relaxing the requirement of asymptotic Minkowski space, one can set $B_1=1+\frac{2 GM}{\sqrt{\pi } \sqrt{\theta }}$ and $B_0=0$ in Eq.(\ref{fr1}). The NCBH obtained in this way is, in fact, a regular black hole.
In this case, the metric function is
\be
f(r)=1+\frac{2 GM}{\sqrt{\pi } \sqrt{\theta }}-\d{2GM}{r\sqrt{\pi}} \gamma(1/2,r^2/4\theta)=1+\frac{GM r^2}{6 \sqrt{\pi } \theta ^{3/2}}+O(r^4).
\ee
Because $M,~\theta$ are both positive, in short distance there is a anti-de Sitter (not de Sitter !) core. The effective cosmological constant is $\Lambda_{eff}=-\frac{GM }{2 \sqrt{\pi } \theta ^{3/2}}=-4\pi G\rho(0)$. At the origin the Ricci scalar is $R=4\Lambda_{eff}$.

(3). For the metric ansatz, Eq.(\ref{metric1}), the equation of state naturally satisfies $\rho(r)=-p_r(r)$  due to $R^t_{~t}=R^r_{~r}$. Thus, we have two independent equations with two unknown functions $f(r)$ and $p_\perp(r)$. There is no room to add other requirement for the equation of state. In fact, to obtain an asymptotic Minkowskian, regular black hole, $p_r(r)$ and $p_\perp(r)$ should be asymptotically vanishing at infinity and finite at the horizon and at the origin\cite{Nicolini.015010.2010}. So, if considering a more general metric ansatz, can we find a reasonable regular NCBH in Rastall gravity? Below we will do this work.

Now we take the more general metric ansatz:
\be\label{metric2}
ds^2 =-e^{A(r)}B(r) dt^2 +\frac{dr^2}{B(r)} +r^2
d\Omega^2.
\ee
Using the field equation of Rastall gravity, Eq.(\ref{eom}), we have
\bea
G^t_{~t}&=&\frac{r B'+B-1}{r^2}=4\pi G \left(-\rho-\d{1}{2}T\right), \label{G00}\\
G^r_{~r}&=&\frac{2 r B A'+r B'+B-1}{r^2}=4\pi G \left(p_r-\d{1}{2}T\right),\label{G11}\\
G^\theta_{~\theta}&=&G^\phi_{~\phi}=\frac{\left(3 r A'+2\right) B'+2 B \left(r A''+r A'^2+A'\right)+r B''}{2 r}=4\pi G \left(p_\perp-\d{1}{2}T\right),\label{G22}
\eea
where $T=2p_\perp+p_r-\rho$ is the trace of the energy-momentum tensor.
According to Eq.(\ref{eom2}), one can obtain another equation
\be\label{emeq}
B\left[r \left(2 \rho  A'+p_r'-2p_\perp' +\rho'\right)+2 p_r \left(r A'+2\right)-4 p_\perp\right]+r B' (p_r+\rho)=0.
\ee
It should be noted that Eq.(\ref{emeq}) can also be obtained from the field equations by eliminating the $A''(r)$ and $B''(r)$. Therefore, we have three independent equations, but with four unknown functions: $A(r),B(r),~p_r(r),~p_\perp(r)$.
To solve these equations, we have to add an extra requirement on equation of state of the matter. For simplicity, we assume
\be\label{Trho}
T=2\rho, \quad \text{namely}, \quad 2p_\perp+p_r=3\rho.
\ee
In this case, Eq.(\ref{G00}) is just the $(^t_t)$ component of Einstein field equations, from which we can derive the metric function $B(r)$
\be\label{sol2}
B(r)=1-\d{4GM\gamma(3/2,r^2/4\theta)}{r\sqrt{\pi}},
\ee
where $\gamma(3/2,r^2/4\theta)$ is the lower incomplete gamma function with definition
  \be
  \gamma(3/2,r^2/4\theta)=\int_0^{r^2/4\theta}dt~ t^{1/2}e^{-t}.
  \ee
Clearly, The solution in Eq.(\ref{sol2}) is just the NCSBH in GR\cite{Nicolini.547.2006}. However, in our metric ansatz there is an extra $e^{A(r)}$. Except for $e^{A(r)}=1$, our black hole solution is different from the NCSBH in GR. 
  
\begin{figure}[htp]
\center{
\includegraphics[width=7cm,keepaspectratio]{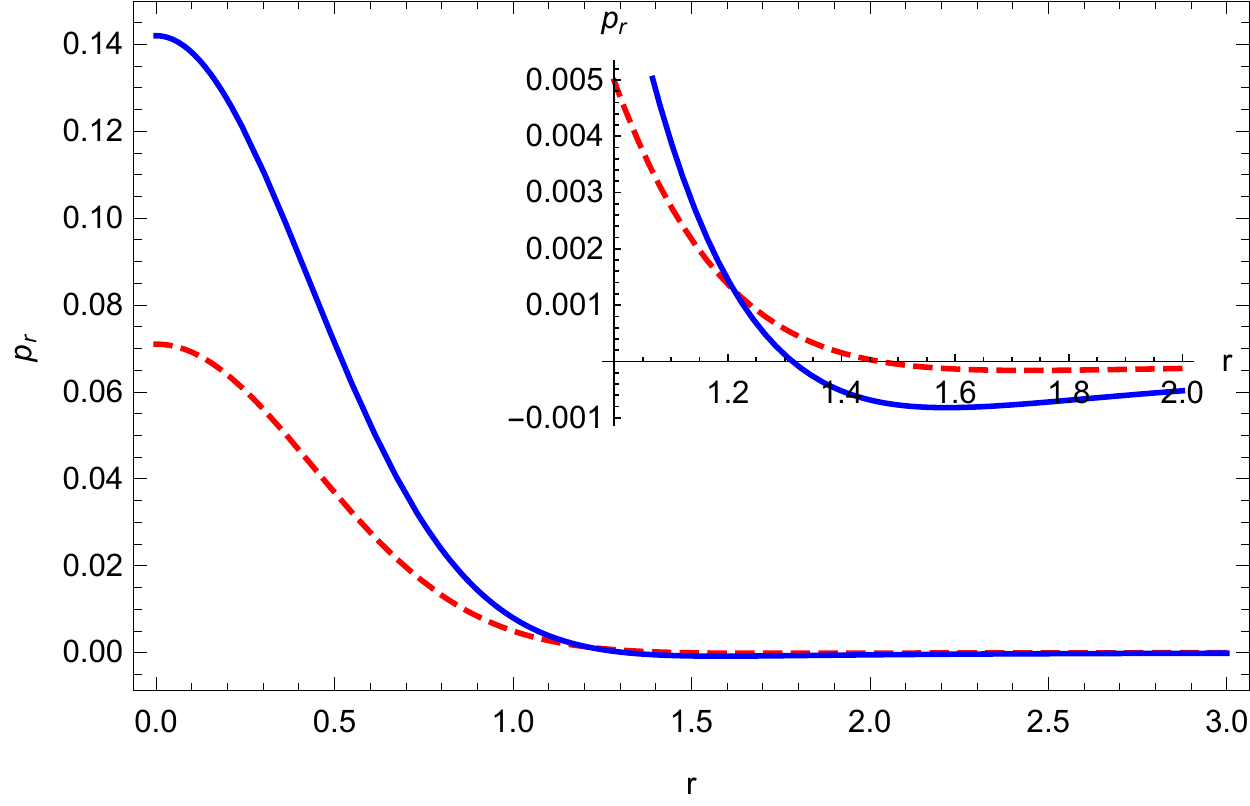}
\caption{$p_r$ as function of $r$. The solid (blue) and dashed (red) curves correspond to the cases with $M=0.2,~\theta=0.1$ and $M=0.1,~\theta=0.1$, respectively. }\label{pr}}
\end{figure}

Combining Eqs.(\ref{G00}),(\ref{G11}) and (\ref{emeq}), we can obtain a first-order differential equation
on $p_r(r)$,
\be
2 r B \left(p_r'-\rho '\right)+(5 B+1) p_r+(1-7 B) \rho +4 \pi  G r^2 p_r^2-4 \pi  G r^2 \rho^2=0.
\ee
This equation can be numerically solved subject to appropriate boundary conditions, which are $p_r(0)=\text{finite}$ and $p_r(\infty)=0$.
In Fig.\ref{pr}, we show the behaviors of the radial pressure $p_r(r)$ for $G=1$ and different values of $M$ and $\theta$. It is shown that $p_r$ is asymptotically flat near the origin and in large distance.
$p_r$ is positive for small $r$. It becomes negative at some point and tends to zero at infinity.

\begin{figure}[htp]
\center{
\includegraphics[width=7cm,keepaspectratio]{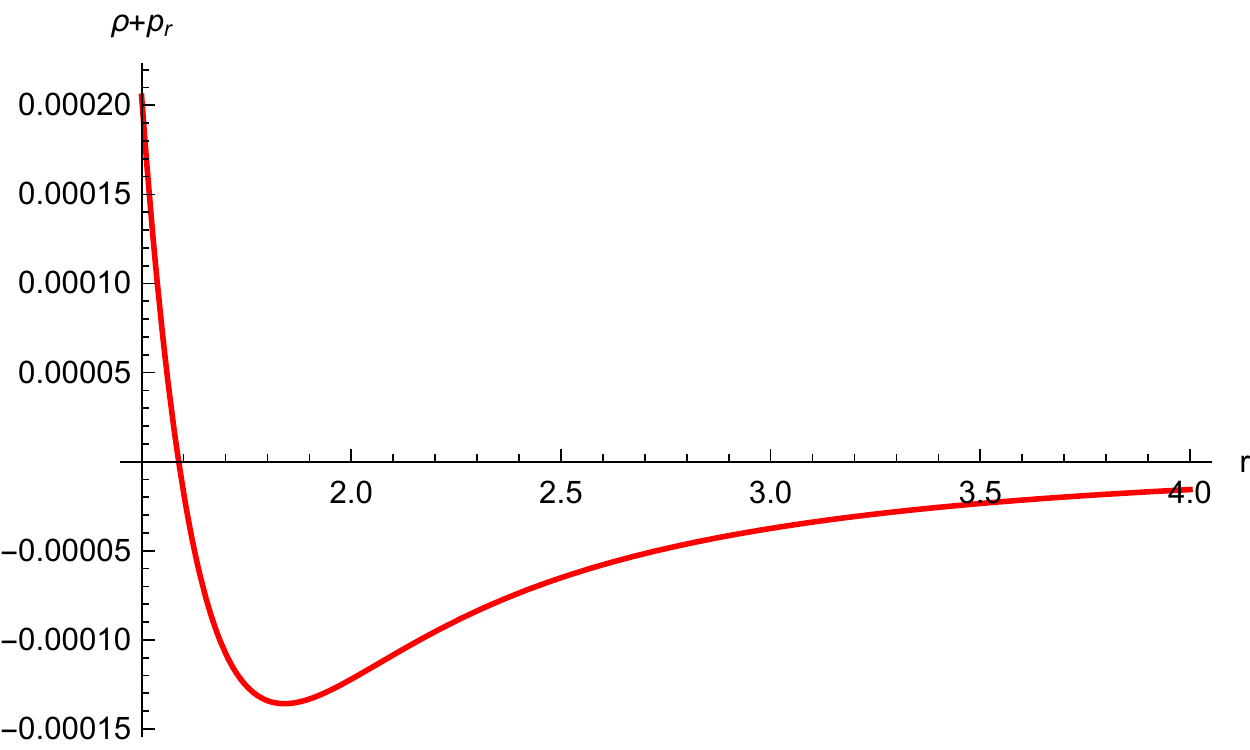}
\caption{$\rho+p_r$ as function of $r$ with $M=0.1,~\theta=0.1$. }\label{rhopr}}
\end{figure}

The behavior of $p_\perp(r)$ is similar to that of $p_r(r)$ except that its value is always positive. Thus, $\rho+p_\perp$ is always positive. Due to Eq.(\ref{Trho}), $\rho+p_r+2p_\perp=4\rho$ is also always positive. However, as is shown in Fig.\ref{rhopr}, $\rho+p_r$ is negative when $r$ is larger than some value $r=r_0$. Therefore, the weak and strong energy conditions are both violated when $r>r_0$. They are fulfilled only in the region $r<r_0$.

\begin{figure}[htp]
\center{
\includegraphics[width=7cm,keepaspectratio]{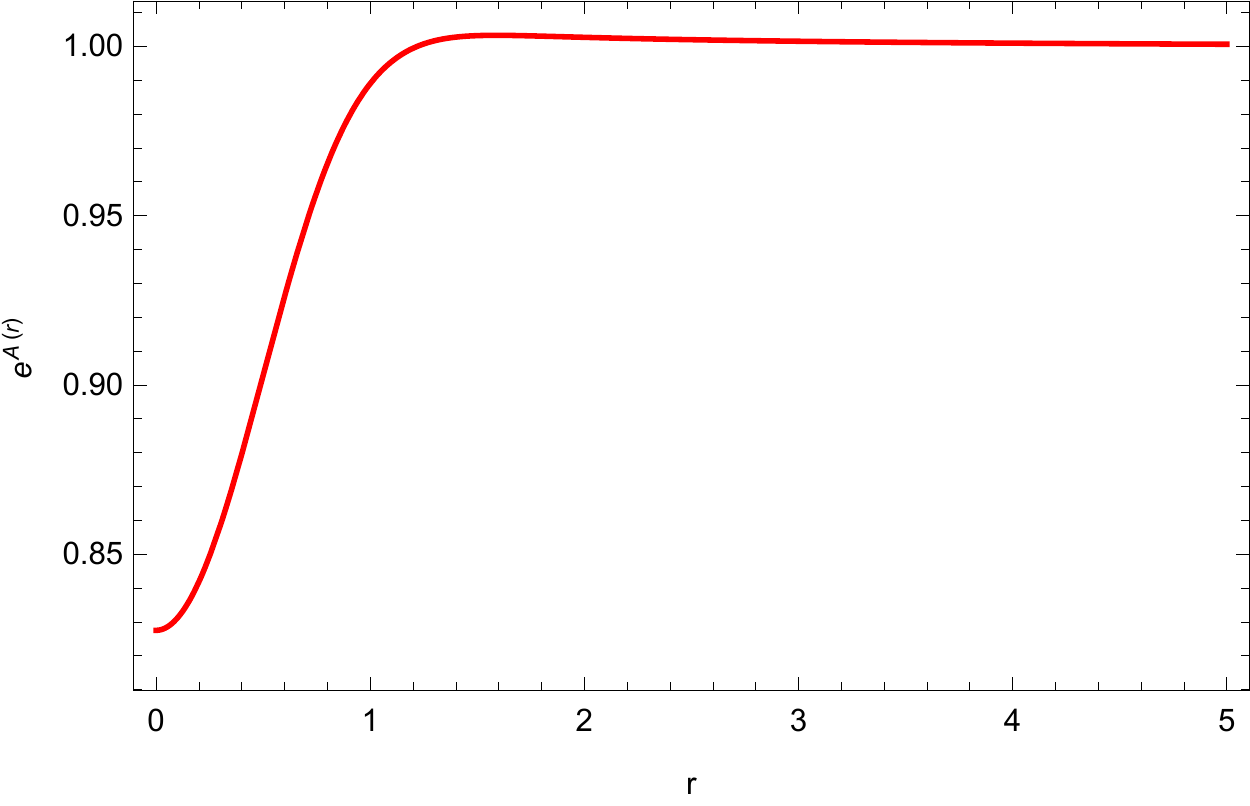}
\caption{$e^{A(r)}$ as function of $r$ with $M=0.1,~\theta=0.1$. }\label{Ar}}
\end{figure}

The behavior of the function $e^{A(r)}$ has been shown in Fig.\ref{Ar}. It can be seen that $A(r)$ approaches to zero at infinity, thus $e^{A(r)}$ tends to one. This assures that the black hole solution is asymptotically Minkowskian. At the origin $A(r)$ has a finite, negative value. These behaviors of $A(r)$ are insensitive to the values of $M,~\theta$. The temperature of this NCBH is
\be
T_H=\d{1}{4\pi r_{H}}e^{A(r_{H})}\left[1-\d{r_{H}^3}{4\theta^{3/2}}\d{e^{-r_H^2/4\theta}}{\gamma(3/2;r_H^2/4\theta)}\right],
\ee
which has an extra factor $e^{A(r_{H})}$ compared to the temperature of NCSBH in GR.  In fact, the effect of $e^{A(r_H)}$ on the temperature can be negligible at larger $r$ due to the behavior of $e^{A(r)}$ depicted in Fig.\ref{Ar}. For very small $r$, it seems that the factor $e^{A(r_H)}$ will lower the value of the temperature. However, for smaller $r$, the temperature of the NSBH becomes negative and thus is physically unacceptable.

\section{Conclusion and Discussion}
\label{Conclusions}

In this paper we studied the noncommutative geometry inspired black hole solutions in Rastall gravity. 
For comparison, we first solve the static, spherically symmetric solution in Rastall gravity with the point-like source. As is shown in Eq.(\ref{noNC}), there are two integration constant $C_1$ and $C_0$ in the metric function. If requiring the spacetime is asymptotically Minkowskian, we also found Schwarzschild solution in Rastall gravity.

Next, we considered the influences of the noncommutativity of spacetime and introduced the Gaussian-distribution smeared source. Considering the special metric ansatz: $g_{00}=g_{11}^{-1}$, we obtained the NCSBH in Rastall gravity. As is well-known,
NCSBH in GR is a kind of regular black hole. The singularity at the origin is replaced by a de Sitter core, which leads to finite curvature invariants at the origin.
Analogous to its counterpart in GR, the NCSBH in Rastall gravity also approaches to the Schwarzschild black hole at infinity. But unlike the NCSBH in GR, it does not tends to the de Sitter space near the origin. The absence of the de Sitter core in short distance makes the NCSBH in Rastall gravity not be a regular black hole. Clearly, although the noncommutativity must lead to the avoidance of curvature singularity of black holes in GR, it is not true for the modified gravities.

For different values of $M/\sqrt{\theta}$, the NCSBH in Rastall gravity can have zero or one event horizon. So its geometric structure is different from those of the Schwarzschild black hole and the NCSBH in GR. We also analyzed the temperature of the black hole. We found that it has the similar behavior to that of NCSBH in GR for large $r_H$. However, the NCSBH in Rastall gravity will leave behind a point-like massive remnant as the temperature tends to zero, while the NCSBH in GR has a zero-temperature remnant with finite radius at last.

By considering a more general metric ansatz with $g_{00}\neq g_{11}^{-1}$, we found another NCBH in Rastall gravity, which is indeed a regular black hole. Although there is an extra factor $e^{A(r)}$ in the metric, the geometric structure and thermodynamic behavior are similar to that of NCSBH in GR, because the value of $e^{A(r)}$ is close to one on a large scale. However, the radial pressure and tangential pressure are different from those of NCSBH in GR due to the special equation of state of the matter we assumed.

In this work we only considered the special case with the Rastall parameter satisfying $\kappa \lambda=1/2$. It is also of great interest to extend our current study to other choices of the Rastall parameter.
To further understand NCBHs, one can even bring in the noncommutativity in other modified gravities. The Gaussian-distribution smeared source we used, is just one possible matter source. In fact, considering the noncommutativity of spcetime, there are several other possible choices\cite{KN.2008,SK.2017}. Even without considering the noncommutativity, but only a kind of special stress-energy tensor, one can also construct nonsingular black holes in GR\cite{Dymnikova.235.1992}. With these matter sources, one can also study black holes in modified gravities. All these will be left as further possible directions.

\acknowledgments
This work is supported in part by the National Natural Science Foundation
of China (Grant Nos.11605107, 11475108) and by the Natural Science Foundation of Shanxi (Grant No.201601D021022).

\bibliographystyle{JHEP}
%\bibliography{H:/mms/References/Rastallgravity}

\end{document}